\documentclass[11pt]{article}
\usepackage{graphicx}
\usepackage{amsmath}
\usepackage{amssymb}
\usepackage{epsfig}
\usepackage[latin1]{inputenc}
\usepackage{color}
\usepackage[latin1]{inputenc}
\usepackage{epsfig}
\usepackage{color}
\usepackage{graphics}
\usepackage{graphicx}

\usepackage{amsthm}
\usepackage{amssymb}
\usepackage{amsmath}
\usepackage{amsfonts}

\numberwithin{equation}{section}
\newcommand{\beq}{\begin{equation}}
\newcommand{\eeq}{\end{equation}}
\newcommand{\ber}{\begin{eqnarray}}
\newcommand{\eer}{\end{eqnarray}}
\newcommand{\bea}{\begin{align}}
\newcommand{\ea}{\end{align}}
\newcommand{\bes}{\begin{split}}
\newcommand{\es}{\end{split}}

\def\<{\langle}
\def\>{\rangle}

\def\s{\sigma}

\def\om{\omega}

\def\<{\langle}
\def\>{\rangle}

\makeatletter
\begin{document}
\baselineskip=15.5pt \pagestyle{plain} \setcounter{page}{1}

\vskip 1.7 cm

\title{Three-point functions in superstring theory on AdS$_{3}\times$S$^3
\times$T$^{4}$ }

\vskip 1.5cm
\author{Carlos A. Cardona  $^{1,}$ \footnote{Email: cargicar@iafe.uba.ar} ~~ 
and
  ~~ Carmen A.
N\'u\~nez ${}^{1,~ 2,}$ \footnote{Email: carmen@iafe.uba.ar}}

\vskip 1.2cm
\date{
\small ${}^1$ Instituto de Astronom\'{\i}a y F\'{\i}sica del Espacio
  (CONICET-UBA).\\
C.~C.~67 - Suc.~28, 1428 Buenos Aires, Argentina \\ and \\
${}^2$   Departamento de F\'\i sica,
FCEN, Universidad de Buenos Aires. \\ Ciudad Universitaria, Pab. I,
  1428 Buenos Aires,
Argentina.}

\maketitle

\vskip 0.5cm

\begin{abstract}
We consider   R and NS spectral flow sectors of type IIB superstring theory
on AdS$_3\times$ S$^3\times $T$^4$ in the context of the
AdS$_3$/CFT$_2$ correspondence.
We present a derivation of the vertex operators creating spectral 
flow images of chiral
primary
states previously proposed in the literature.
We compute  spectral flow conserving
three-point functions involving these operators on the sphere. Using the 
bulk-to-boundary dictionary, we compare the results with the corresponding
correlators in the dual conformal field theory, the
 symmetric product orbifold of T$^4$. In the limit of small string
 coupling, agreement is found in all the cases considered.

\vspace{1cm}

\end{abstract}

\section{Introduction}

A systematic understanding of the duality between type IIB
 superstring theory on
AdS$_3\times$ S$^3\times $T$^4$ and
 the ${\cal N}=(4,4)$ non-linear sigma model
 on the moduli space of Yang-Mills instantons on T$^4$
has been achieved along recent years,
 based on  early work in
 \cite{malda}-\cite{ david}.

The instanton moduli space is a
deformation of the symmetric product of N copies of T$^4$, namely
$Sym({\rm T}^4)^{\rm N}\equiv({\rm T}^4)^{\rm N}/S_{\rm N}$ 
\cite{Douglas:1995bn} and the worldsheet of the superstring 
is an ${\cal N}=1$ SL(2,$\mathbb R)\times$ SU(2) WZNW model. 
 In the large N limit,
 twisted
states in $Sym({\rm T}^4)$  map to single states of short strings
 \cite{Gaberdiel, Pakman} described by
discrete representations of SL$(2,\mathbb R)\times$ SU$(2)$ and their
spectral flow images \cite{Maldacena:2000}.
Agreement between the  spectrum and  three-point functions of
unflowed chiral primary string states and
 the corresponding dual counterparts was found in
\cite{Gaberdiel, Pakman, ps}.
Conversely,  the non-trivial spectral
flow sectors of the string theory  have been less studied and present
some unclear features, such as the apparent lack of
certain  string states
 in the spectrum of the superconformal field theory (SCFT)
 \cite{Argurio} and various technical difficulties
in the computation of correlation functions. Some preliminary
results were obtained in \cite{Giribet:2007wp} where, in particular,
a bulk-to-boundary dictionary for 1/2 BPS states in the flowed sectors was
proposed. 

The aim of this paper is to 
study  this holographic map by exploring three-point functions in
both sides of the duality.
The computation of worldsheet correlators basically involves 
three parts reflecting the fact that 
the theory   is a direct product of  free fermions and bosonic
SU(2) and SL(2,$\mathbb R)$ WZNW models.
 The relevant three-point functions of the free fermions
 and SU(2) bosons were obtained in \cite{Giribet:2007wp}. Here we
evaluate spectral flow conserving
three-point
functions on the sphere 
involving spectral flow images of chiral primary  string
states in the Neveu-Schwarz (NS) and Ramond (R) sectors of the SL(2,
$\mathbb R)$ WZNW model in order
to complete the construction of these amplitudes in the full string theory and 
compare them
with the conjectured dual correlators in the symmetric orbifold of
T$^4$
 obtained in \cite{Jevicki,
  Lunin, Lunin2}. 
Our results  confirm the
 agreement of the string amplitudes
with the corresponding counterparts in the dual theory.

The paper is organized as follows. After setting the notations in the next 
section, in
section 3
we present a derivation of the vertex operators
creating spectral flow images of chiral primary
string states in NS and R sectors which were  proposed in
\cite{Giribet:2007wp}.
In section 4 we compute the SL(2, $\mathbb R$) part of the
spectral flow conserving three-point
functions involving these chiral operators 
and, after adding the fermionic and SU(2) parts computed in 
\cite{Giribet:2007wp},
 we compare our results with the conjectured corresponding correlators
in the dual SCFT.
Finally, section
5 contains the conclusions. In the Appendix
we compute the Clebsch-Gordan coefficients needed to
construct 
vertex operators in  product representations of SL$(2,\mathbb R)$.

\section{Notations}

In order to set the notations in this section
we briefly review basic aspects of the dual theories.

\medskip

\noindent{\bf 2.1} {\it Review of
type IIB superstring on} AdS$_3\times$ S$^3\times$ T$^4$

\medskip

Type IIB superstring theory on AdS$_3\times$ S$^3\times$ T$^4$
was originally studied in \cite{Kutasov98, Kutasov:1999, aharony, david,
 Argurio, kll}. It has
   $\widehat{{\rm SL}(2)}\times \widehat{{\rm SU}(2)}\times
\widehat{{\rm U}(1)}^4$
affine worldsheet symmetry which allows to perform explicit calculations.
The $\widehat{{\rm SL}(2)}$ and $\widehat{{\rm SU}(2)}$ supercurrents
$\psi^A+\theta J^A$  and $\chi^A+\theta K^A$, respectively,
satisfy the following OPE
\begin{eqnarray}
J^A(z)J^B(w) \sim  \frac{\frac {k}2 \eta^{AB}}{(z-w)^2}+
\frac{i\epsilon^{AB}_{~~~C}J^C(w)}{z-w}\, ,&& K^A(z)K^B(w) \sim
\frac{\frac {k}2 \delta^{AB}}{(z-w)^2}+
\frac{i\epsilon^{AB}_{~~C}K^C(w)}{z-w}\, ,
\nonumber\\
J^A(z)\psi^B(w) \sim
\frac{i\epsilon^{AB}_{~~~C}\psi^C(w)}{z-w}\, ,&& K^A(z)\chi^B(w) \sim
\frac{i\epsilon^{AB}_{~~~C}\chi^C(w)}{z-w}\, ,
\nonumber\\
\psi^A(z)\psi^B(w) \sim  \frac{\frac {k}2 \eta^{AB}}{z-w}\, ,&&
\chi^A(z)\chi^B(w) \sim  \frac{\frac {k}2 \delta^{AB}}{z-w}\, ,\label{algebras}
\end{eqnarray}
 with $A=0,1,2$, $\epsilon^{012}=1$ and
$\eta^{AB}=(-,+,+)$.
It is convenient to introduce new currents as
\begin{eqnarray}
J^A(z)=j^A(z)+\hat{j}^A(z)\, ,\qquad
K^A(z)=k^A(z)+\hat{k}^A(z)\, ,\label{split}
\end{eqnarray}
where  $j^A$ ($\hat j^A$) and $k^A$ ($\hat k^A$)
 generate SL(2)$_{k+2}$ (SL(2)$_{-2}$) and SU(2)$_{k-2}$ (SU(2)$_2$)
affine algebras, respectively, with
\begin{eqnarray}
\hat{j}^A(z)=-\frac i{k} \epsilon^A_{~~BC} \psi^B(z)\psi^C(z) ,\qquad
\hat{k}^A(z)=-\frac {i}{k} \epsilon^A_{~BC} \chi^B(z)\chi^C(z) .
\end{eqnarray}
The $\widehat {{\rm U}(1)}^4$ is realized in terms of free bosonic currents
$i\partial Y^i$ and free fermions $\lambda ^i, i=1, 2,3,4$.

The stress tensor and supercurrent are given by
\begin{eqnarray}
T(z)&=&\frac{\eta_{AB}}{k}(j^Aj^B -\psi^A \partial \psi^B)+
\frac {\delta_{AB}}k(k^Ak^B-\chi^A\partial\chi^B) +\frac 12(\partial Y^i\partial
Y_i-\frac 12\lambda^i\partial \lambda_i),\nonumber\\
G(z)&=& \frac{2}{k} ( \eta_{AB}\psi^A j^B + \frac{2i}{k}
\psi^0\psi^1\psi^2)
+\frac 2k (\delta_{AB}\chi^A k^B -\frac {2i}{k}\chi^0\chi^1\chi^2)
+\lambda^i\partial Y_i\, .
\label{supercurrent}
\end{eqnarray}

The spectrum of the theory is built from those of the
SL(2,$\mathbb R)$ and SU(2) WZNW models.
The Hilbert space of the former  \cite{Maldacena:2000} is decomposed
into unitary
representations of the SL(2,$\mathbb R)\times$ SL(2,$\mathbb R)$ current
algebra \footnote{Actually, the spectrum is built on representations of
the universal cover of SL(2, $\mathbb R)$, to which we refer simply as
SL(2,$\mathbb R)$ for short.}, namely the discrete lowest- and highest-weight
representations ${\cal D}_h^{\pm}\otimes{\cal
D}_h^{\pm}$ with
$h\in\mathbb R$, $\frac 12<h<\frac{k+1}2$ and $m=\pm h, \pm h\pm 1,
\dots$, the
continuous representations ${\cal C}_h^{\alpha}\otimes{\cal
  C}_h^{\alpha}$ with
$h=\frac 12+i\mathbb R$, $m=\alpha+\mathbb Z$, $\alpha\in [0,1)$,
their current algebra descendants and spectral flow
  images,  $\widehat{\cal D}_h^{\pm,w}
\otimes\widehat{\cal D}_h^{\pm,w}
, \widehat{\cal C}_h^{\alpha,
    w}\otimes\widehat{\cal C}_h^{\alpha,
    w}$
 with $w\in\mathbb Z$ and the same spin and  amount of spectral flow on the
 left- and right-moving sectors.
Primary operators of spin $h$
and
worldsheet
conformal dimension
$\Delta^{sl}(\Phi_{h})=-\frac{h(h-1)}{k}$, satisfy
\begin{equation}
j^a(z)\Phi_h(x,\overline x; w,\overline w)\sim \frac{D^a_x
\Phi_h(x,\overline x;w,\overline
  w)}{z-w}\, , \quad a=0, \pm\, ,
\end{equation}
where
$D^+_x=\partial_x ,\, D^0_x=x\partial_x+h$ and $
D^-_x=x^2\partial_x+2hx$.
Expanding in modes as
\begin{equation}
\Phi_h(x,\overline x)=\sum_{m,\overline m}\Phi_{h,m,\overline
  m}x^{-h+m}\overline x^{-h+\overline m}\, ,
\end{equation}
one can read the action of the zero modes of the currents on
$\Phi_{h,m,\overline m}$, namely
\begin{eqnarray}
j_0^0\Phi_{h,m,\overline m}= m\Phi_{h,m,\overline m}\, ,\qquad
j_0^\pm\Phi_{h,m,\overline m}= \left [m\mp(h-1)\right ]\Phi_{h,m\pm1,\overline
  m},\quad (m\ne \pm h)\, ,
\label{modesuf}
\end{eqnarray}
and $j_0^-\Phi_{h,h,\overline m}=j_0^+\Phi_{h,-h,\overline m}=0$.

Similarly, the primary fields of the SU(2)$_{k-2}$ WZNW model with
conformal dimension
$\Delta^{su}(V_j)=\frac{j(j+1)}k$
verify
\begin{equation}
k^a(z)V_j(y,\overline y;w,\overline
w)\sim \frac{P^a_yV_j(y,\overline y; w,\overline w)}{z-w}\, ,
\end{equation}
with
$P^+_y=\partial_y,\, P^0_y=y\partial_y-j$,
$P^-_y=-y^2\partial_y+2jy$ and can be expanded in modes as
\begin{equation}
V_j(y,\overline y)=\sum_{m',\overline m'=-j}^jV_{j,m',\overline m'}
y^{j+m'}\overline
y^{j+\overline m'}\, .
\end{equation}
The spin $j\in \mathbb Z/2$ is bounded by
$0\le j\le \frac{k-2}2$ and
$k_0^+V_{j,j,\overline m'}=k_0^-V_{j,-j,\overline m'}=0$,
\begin{eqnarray}
k_0^0V_{j,m',\overline m'}=m'V_{j,m',\overline m'}\, ,\qquad
k_0^\pm V_{j,m',\overline m'}=(\pm m'+1+j)V_{j,m'\pm 1,\overline m'}\,
,\quad (m'\ne\pm j)\, .
\end{eqnarray}

In the fermionic sector, the fields
$\psi^a$ transform in the
spin $\hat h=-1$
representation of the global
SL(2,$\mathbb R)_{-2}$ algebra and
$\chi^a$
transform in the spin $\hat j=1$ of the SU(2)$_2$ global algebra.

Vertex operators creating unflowed physical states
in the NS sector
were constructed in  \cite{kll}. For short, we display
only the holomorphic indices.
The chiral (antichiral) primaries satisfy the condition ${\cal
  H}={\cal J}$
(${\cal H}=-{\cal J}$),
 ${\cal H}$ being the spacetime
conformal dimension and ${\cal J}$  the  SU(2) charge, which
implies $h=j+1$. In the $-1$ picture,
they are given by
\begin{eqnarray}
{\cal W}_{h,m, m'} & = &
e^{-\varphi}(\psi\Phi_{h,m})_{h- 1,m^{}_T}
V_{j,m'}\, ,\label{vfsl}\\
{\cal Y}_{h,m,, m'} & = & e^{-\varphi}
\Phi_{h,m}(\chi V_{j,m'})_{h, m'_T}\, ,\label{vfsu}
\end{eqnarray}
where $(\psi\Phi_{h,m})$ and $(\chi V_{j,m'})$ denote the product
representations of $J^a$ and $K^a$, respectively,
$m^{}_T=h-1,h,h+1,\cdots$ and $m'_T=-h,-h+1,\cdots, h$.

To study the Ramond sector one needs to construct the spin fields for
$\psi^a, \chi^a, \lambda^i$ \cite{kll}. 
It is convenient to have a bosonized form
of the fermions such as
\begin{eqnarray}
\partial H_1=\frac 2ki\psi^+\psi^-,\quad \partial H_2=\frac
2ki\chi^+\chi^-,\quad \partial H_3=-\frac 2k i\psi^3\chi^3,\quad
\partial H_4=\lambda^1\lambda^2, \quad \partial
H_5=\lambda^3\lambda^4\, .
\end{eqnarray}
The spin fields take the form 
$S_{[\epsilon_1,\cdots,\epsilon_5]}={\rm exp}\frac
i2\sum_{i=1}^5\epsilon_iH_i$, with $\epsilon_i=\pm 1$. 
They transform as two
copies of ($\frac 12,\frac 12$) under SL(2)$\times$ SU(2).
GSO projection 
requires $\prod_{i=1}^5\epsilon_i=+1$ and 
BRST invariance demands $\prod_{i=1}^3\epsilon_i=-1$. 
Following \cite{Giribet:2007wp}
we define the  spin fields associated with
$\psi^a, \chi^a$ as $\tilde
S_{[\epsilon_1,\epsilon_2,\epsilon_3]}={\rm exp}\frac
i2(\epsilon_1H_1+\epsilon_2H_2+\epsilon_3H_3)$. 

Decomposing
 the
product $(\tilde S\Phi_{h,m} V_{j,m'})$ into representations of the
total currents $J^a, K^a$, the chiral vertex operators in the $-\frac 12$
picture take the form
\begin{equation}
{\cal R}^\pm_{h,m, m'}=e^{-\frac \varphi 2} (\tilde S \Phi_{h,m}
V_{j,m'})_{h-\frac
  12,
m^{}_T+\frac 12; j+\frac 12, m'_T+\frac 12}e^{\pm i(\hat H_4-\hat H_5)}\, ,
\end{equation}
where $H_i$ are redefined as $\hat H_i=H_i+\pi\sum_{j<i}N_j,
N_j=i\oint\partial H_i$.

\medskip

\noindent{$\bullet$}{\it ~~Spectral flow}

\medskip

The
algebras
(\ref{algebras})
are invariant under the following spectral flow automorphisms
\begin{eqnarray}\label{slspectralflow}
\widetilde J^0_n={J}^0_n-\frac{k}{2}w\delta_{n,0},\,~~
\widetilde J^{\pm}_n={J}^{\pm}_{n\pm w}, &&
\widetilde K^0_n= K^0_n+\frac {k}2 w'\delta_{n,0}, \,~~
\widetilde K^\pm_n= K^\pm_{n\pm w'}\, .\nonumber
\end{eqnarray}
The currents
$j^a$, $\hat{j}^a$, $k^a$ and $\hat k^a$
transform under
spectral flow as
\begin{eqnarray}
j^0_n=\tilde{j}^0_n+\frac{k+2}{2} w\delta_{n,0}\, , ~~
 j^{\pm}_n=\tilde{j}^{\pm}_{n\mp w}\, , && k^0_n=\tilde
 k^0_n-\frac{k-2}2w'\delta_{n,0}\, , ~~ k^\pm_n=\tilde k^\pm_{n\mp w'}\, ,\\
\hat{j}^0_n = \tilde{\hat{j}}^0_n-w\delta_{n,0},\, ~~
 \hat{j}^{\pm}_n=\tilde{\hat{j}}^\pm_{n\mp w} \, ,  && \hat
 k^0_n=\tilde{\hat k}^0_n-w'\delta_{n,0}\, ,~~\hat
 k^\pm_n=\tilde{\hat k}^\pm_{n\mp w'}\, ,
\label{spf}
\end{eqnarray}
and
the modes of the Virasoro generators,
$L^{sl}_n=l^{sl}_n+\hat{l}^{sl}_n,
 L^{su}_n=l^{su}_n+\hat l^{su}_n$, as
\begin{eqnarray}
\tilde L^{sl}_n={L}^{sl}_n+w\tilde{J}^0_n+\frac{k}{4}w^2\delta_{n,0}\, ,
&& \tilde L^{su}_n= L^{su}_n+w'\tilde K^0_n-\frac k4 w^{\prime
  2}\delta_{n,0}\,
\nonumber\\
l^{sl}_n=\tilde{l}^{sl}_n-w\tilde{j}^0_n-\frac{k+2}{4}w^2\delta_{n,0}\,
,&&
l^{su}_n=\tilde l^{su}_n-w'\tilde k^0_n+\frac{k-2}4 w^{\prime 2}\, ,\nonumber\\
\hat{l}^{sl}_n=\tilde{\hat{l}}^{sl}_n-w\tilde{\hat{j}}^0_n+\frac{1}{2}w^2\delta_{n,0}\,
,
&&\hat{l}^{su}_n=\tilde{\hat{l}}^{su}_n-w'\tilde{\hat{k}}^0_n+\frac{1}{2}
w^{\prime
  2}
\delta_{n,0}\, .
\label{virsplit}
\end{eqnarray}
The closure of
 the
SL(2,$\mathbb R)$ and SU(2) algebras requires the
 same amount of spectral flow $w$ ($w'$) for $j^a$ and $\hat j^a$ ($k^a$ and
$\hat k^a$).
The spectral flow maps 
primaries to descendants of
SU(2) and it
generates new representations in SL(2,$\mathbb R)$ 
\cite{Maldacena:2000}. For the sake of simplicity, we restrict
to  $w>0$ in this
section.

To construct  spectral flow images of chiral primaries in generic
frames,
we consider the SL(2,$\mathbb R)$ sector first.
A $w=0$ affine primary   is mapped by
the spectral flow to a lowest-weight
state of the global algebra $\Phi^{h,w}_{ H, M}$ with $H = M$
satisfying \cite{Maldacena:2000}
\ber\label{j3flow}
j^0_0\Phi^{h,w}_{ H, M}
&=&M\Phi^{h,w}_{ H, M}= \left({m}+\frac{k+2}{2}w\right)
\Phi^{h,w}_{ H, M}
\, ,\label{lowest}\\
l_0\Phi^{h,w}_{ H, M}&=&\left(-\frac{{h}({h}-1)}{k}-w{m}-
\frac{k+2}{4}w^2\right)
\Phi^{h,w}_{ H, M}\, . \label{conforworll}
\eer

In the fermionic SL(2,$\mathbb R$) sector, an interesting description 
of the spectral flow 
was presented by A. Pakman in  \cite{pak}.
Using (\ref{spf}) $-$
(\ref{virsplit}), the fermions $\psi^a$
in the spectral flow frame  obey
\ber
\hat{\jmath}^0_0\psi^{a}=(a-w)\psi^{a} ,\qquad
\hat{\jmath}^-_0\psi^a=
\tilde{\hat{\jmath}}^-_w\psi^a=0\, ,\qquad
\hat{l}_0\psi^a=
(\frac{1}{2}-wa+\frac{1}{2}w^2)\psi^a\, ,
\label{conforworlhat}\eer
$i.e.~ \psi^a$ is a lowest-weight field with angular momentum $\hat
h=a-w$.
Acting with  $\hat j^+_0$, one obtains the global
representation in the $w$ sector as
$\psi^{|\hat h|}_{\hat m}
\sim (\hat{\jmath}^+_0)^n\psi^a$ with
$\hat m=-\hat {h}, \cdots , \hat {h}$  
up to a normalization.

 All   these ingredients allow to construct the  representations of
 $J^a$. We denote
the fields
of the product
  representation in the NS sector
 as $(\psi^{|\hat h|}_{\hat m}\Phi_{H,M}^{w,h})^{}_{{\cal H},
{\cal M}}(z, \overline z)$, where
$
|H-\hat{h}|\leq {\cal H}\leq H+\hat{h}$,
${\cal M}={\cal H}, {\cal H}+1,...$ and their worldsheet conformal weight is
given by
\begin{equation}
\Delta^{sl}\left [(\psi^{|\hat h|}_{\hat m}\Phi_{H,M}^{h,w})^{}_{{\cal H},{\cal M}}
\right ]=-\frac{{h}(h-1)}{k}-w(m-a)
+\frac{1}{2}-\frac{k}{4}w^2\, .
\end{equation}

Repeating the analysis for
SU(2),  one obtains the product representation
$(\chi^{\hat j}_{\hat m'} V^{j,w'}_{J,M'})^{}_{{\cal J},{\cal M}'}$, with
$|J-\hat\jmath |\le {\cal J}\le J+\hat\jmath $,
$-{\cal J}\le{\cal M}'\le {\cal J}$, $J=m'-\frac
{k-2}2w'$, $\hat{\jmath}=|a-w'|$ and
worldsheet conformal weight
\begin{equation}   \Delta^{su}\left [(\chi^{\hat j}_{\hat m'}
V^{j,w'}_{J,M'})^{}_{{\cal J},{\cal M}'}\right ]=
\frac{{j}({j}+1)}{k}-w'({m}'-a)
+\frac{1}{2}+\frac{k}{4}w^{\prime 2}\, .
\end{equation}

In order to construct chiral states,
we apply the spectral flow operation on
 the chiral primaries
 (\ref{vfsl}) and (\ref{vfsu}).
We notice that the physical and chiral state conditions require to
 simultaneously spectral flow
the SL(2,$\mathbb R)$ and SU(2)
product representations and we obtain
\begin{eqnarray}
\mathcal{W}^{h,w}_{{\cal H}, {\cal M}}&=&
e^{-\varphi} (\psi^{w+1}_{\hat m}\Phi^{h,w}_{H,M})^{}_{{\cal H},{\cal M}}
(\chi^{w'}_{\hat m'} V^{j,w'}_{J,M'})^{}_{{\cal J},{\cal M}'} \,
  ,
\label{vertexfloww}
\end{eqnarray}
\begin{eqnarray}
{\cal Y}^{h,w}_{{\cal H},{\cal M}}
& = & e^{-\varphi}
(\psi^w_{\hat m}\Phi^{h,w}_{H,M})^{}_{{\cal H},{\cal M}}
(\chi^{w'+1}_{\hat m'} V^{j,w'}_{J,M'})^{}_{{\cal J},{\cal M}'}\, ,\label{vfc}
\end{eqnarray}
where $\varphi$ is the bosonization of the $\beta, \gamma$ ghosts,
${\cal M}={\cal H}$ and
 ${\cal M}'=-{\cal J}$.
For
generic level $k$, the physical state condition $(L_0-1){\cal W}=0$
implies $h=j+1$, $w=w'$ and
$m'_T=-m^{}_T$ (see (\ref{vfsl}),(\ref{vfsu})),
and similarly for ${\cal Y}$. Analogously, $G_r{\cal W}=(\tilde G_r-
w\tilde\psi_r^0-w\tilde\chi_r^0){\cal W} =0$ ($G_r{\cal Y}=0$) for $r>0$
requires
$m^{}_T=h-1$
($m_T=h$) \cite{Giribet:2007wp}.
Finally,
chirality
 (or antichirality) demands, for both operators ${\cal W}$ and ${\cal Y}$,
\begin{eqnarray}
{\cal H}=m^{}_T+\frac k2 w=\pm{\cal J} .
\end{eqnarray}

To obtain the spectral flowed
$\frac 12$ BPS operators in the Ramond sector
we need the product representation
$(S^{\hat j}_{\hat m,\hat m'}\Phi^{h,w}_{H,M} V^{j,w'}_{J,M'})$.
The discussion about the fermions
 applies analogously to the spin fields, $i.e.$ from the lowest-weight
 component of the  $\hat h=-\hat j=-|w\pm\frac 12|$ spin representation,
 given by
\begin{equation}
S^{w+\frac 12}_{-w-\frac 12,w+\frac 12}
\equiv e^{-i(w+\frac 12)(\hat H_1+\hat H_2)-
\frac i2\hat H_3}\, ,
\end{equation}
one constructs the global representation acting with $\hat j_0^+, \hat k_0^+$.

The chiral fields in
the $w$ sector  are \cite{Giribet:2007wp}
\begin{equation}
{\cal R}^{\pm,h,w}_{\cal H,\cal M}=e^{-\frac{\varphi} 2}
(S^{w+\frac 12}_{\hat m,\hat m'}
\Phi^{h,w}_{H,M} V^{j,w}_{J,M'})^{}_{{\cal H}, {\cal M}, {\cal J}, {\cal M}'}
e^{\pm \frac i2(\hat H_4-\hat H_5)} \, ,\label{vertexram}
\end{equation}
where $S^{w+\frac 12}_{\hat m,\hat m'}$ has conformal weight $\frac
38+w^2+w$, $\hat h=-w-\frac 12=-\hat j$
and
${\cal H}=h-\frac 12+\frac k2 w={\cal J}$. 

\bigskip

\noindent {\bf 2.2} {\it Sigma Model On The Symmetric Product Orbifold
of T$^4$}

\medskip

Type IIB superstring theory on AdS$_3\times$ S$^3\times$ T$^4$
with RR background is
  conjectured to be dual to
the infrared fixed point theory living on a D1-D5 system compactified
  on T$^4$.
It is convenient to use the S-dual description \cite{ms} in terms of
N$_1$ fundamental strings and N$_5$ NS5-branes.
The target space of the SCFT is identified with the
  singular orbifold (T$^4)^{{\rm N}_1{\rm N}_5}/S$(N$_1$N$_5)$, 
where $S($N$_1$N$_5)$
  denotes the permutation group of N$_1$N$_5$ elements.
It was argued in \cite{larsen} that the symmetric orbifold 
corresponds to the point 
N$_5$=1, N$_1$=N.

The chiral spectrum of the sigma model
is built from that of a single copy of
T$^4$ plus operators in the twisted sectors. Each twisted sector
corresponds to one conjugacy class of
$S$(N), labeled by positive integer partitions of
N, namely
\begin{equation}
 \sum_{l=1}^{{\rm N}} l k_l={\rm N}\, ,
\end{equation}
corresponding to permutations with $k_l$ cycles of length $l$. Chiral operators
describing  single particle states in the string theory side
correspond to single cycle twist operators
\cite{ms, deBoer}.
There is one twist field for each conjugacy class of the
permutation group,
and chiral operators corresponding to chiral states in the
 dual string theory can be constructed as a sum
 over the group orbit, namely
\begin{equation}\mathcal
{O}_n^{\epsilon_n,\overline\epsilon_{\overline n}}=\left[n(\rm{N}-n)!
\rm{N}!\right]^{-1/2}\sum_{h\epsilon
S(\rm{N})}\sigma^{\epsilon_n,\overline
\epsilon_{\overline n}}_{h(1\cdots n)h^{-1}}\,
,
\end{equation}
where $\epsilon_n
=\pm 1, a$ and
$\sigma^{1_n}_{(1\cdots n)}$ is a twist field corresponding to
just one single element of $S(\rm{N})$. The global part of the
${\cal N}=(4, 4)$ superconformal algebra forms the supergroup
SU(1,1$|2)_L \times$ SU(1,1$|2)_R$ and contains the R-symmetry group
SU(2)$_L \times$ SU(2)$_R$,  under which the operator
$\mathcal{O}_{n}^{\epsilon_n,\overline\epsilon_{\overline n}}
(x,\overline{x})$ is a chiral state  in a
unitary representation with angular momentum 
\begin{eqnarray}
{\cal H}_n=\frac{n+\epsilon_n}{2}, 
& & 0\leq
{\cal H}_n\leq\frac{{\rm N}+\epsilon_{n}}{2}\, ,\,
  ~~\epsilon_n =\pm 1, ~~~~ \label{spinPS}\\
~~~{\cal H}_n=\frac{n}{2},~~~
&& ~~~ 0\leq
{\cal H}_n\leq\frac{\rm{N}}{2}
\; , \;
 ~~~ ~~~\epsilon_n=a\, , \label{spinPS1}
\end{eqnarray}
and similarly for $\overline \epsilon_{\overline n}$.
Two- and three-point functions on the sphere for $\epsilon_n, \overline
 \epsilon_{\overline n}=\pm 1$, are given by
 \cite{Lunin, Lunin2} 
\footnote{
Contributions from surfaces with higher genus are suppressed in the large N limit.}
\begin{equation}\label{2ptLunin}
\langle \mathcal{O}^{\epsilon_{n}, \overline
\epsilon_{\overline n}}_{n}(x_1,\overline
x_1)\mathcal{O}^{-\epsilon_{n},-\overline
\epsilon_{\overline n}}_{-n}(x_2,\overline x_2) \rangle =
|x_{12}|^{-4{\cal H}_{n}}\ ,
\end{equation}
\ber
\langle {\cal O}^
{\epsilon_{n_1},{\overline \epsilon}_{{\overline n}_1}}_{n_1}
{\cal O}^{\epsilon_{n_2}
{\overline \epsilon}_{\overline n_2}}_{n_2}
 {\cal O}^{\epsilon_{n_3}{\overline
\epsilon}_{\overline n_3}\dagger}_{n_3}\rangle
=
{\sqrt{\frac{n_1n_2n_3}{\rm{ N}}}}\delta^2 ({
\sum_{i=1}^3 {\cal M}_{n_i}} ) C_{n_1n_2n_3}  \, \prod_{i<j}
|x_{ij}|^{-2{\cal H}_{n_in_j}} \, ,
\label{3pfPS}
\eer
where ${\cal H}_{n_1n_2}={\cal H}_{n_1}+{\cal H}_{n_2}-{\cal
  H}_{n_3}$, etc., $-{\cal H}_n\le {\cal M}_n\le{\cal H}_n$ and
the  coefficients $C_{n_1n_2n_3}$ are defined in terms of the
SU(2) 3j symbols as
\ber
C_{n_1n_2n_3}&=&\frac{|\epsilon_{n_1}n_1+
\epsilon_{n_2}n_2+\epsilon_{n_3}n_3+1|^2}{4
n_1n_2n_3}\nonumber\\
&&\times \left |
\begin{pmatrix}{\cal H}_{n_1}&{\cal H}_{n_2}&{\cal H}_{n_3}\\
{\cal M}_{n_1}&{\cal M}_{n_2}&{\cal M}_{n_3}\end{pmatrix}\right |^2
\left|\frac{{\cal
      H}_{n_1n_2}!{\cal H}_{n_2n_3}!{\cal H}_{n_3n_1}!
(\sum_{i=1}^3{\cal H}_{n_i}+1)!}
{(2{\cal H}_{n_1})!(2{\cal H}_{n_2})!(2{\cal H}_{n_3})!}\right
|\, .
\nonumber
\eer
Using (\ref{spinPS}) and ${\cal M}_{n_i}=\pm{\cal H}_{n_i}$,
the delta function in (\ref{3pfPS}) implies ${\cal H}_{n_in_j}=0$ for certain 
$i,j$. Specifying $n_3=n_1+n_2-1$,
 the non-vanishing
three-point
functions are those with $(\epsilon_{n_1},\epsilon_{n_2},\epsilon_{n_3})
=(-,-,-)$ and $(+,-,+)$ and similarly for
$\overline\epsilon_{\overline n_i}$. In this case, the product in the second 
line  reduces to one.

Two other correlators that will be important below have been
 evaluated in the particular case
$n_3=n_1+n_2-1$ \cite{Jevicki}, namely (we omit the obvious
 coordinate dependence)
\begin{eqnarray}
\left\langle{\cal O}^{a,\overline a}_{n_1}
{\cal O}^{-,-}_{n_2}
{\cal O}^{a',\overline a' \dag}_{n_3}
\right\rangle&=&\frac 1{\sqrt{\rm {N}}}\left (\frac{n_1n_3}{n_2}\right
)^{1/2}\delta^{a,a'}\delta^{\overline a,\overline a'}\, ,\label{r1}\\
\left\langle{\cal O}^{a,\overline a}_{n_1}
{\cal O}^{a',\overline a' }
_{n_2}
{\cal O}^{+,+ \dag}
_{n_3}
\right\rangle&=&\frac 1{\sqrt{\rm {N}}}\left (\frac{n_1n_2}{n_3}\right
)^{1/2}\xi^{a,a'}\xi^{\overline a,\overline a'}
\, ,\quad
\xi^{a,a'}=\xi^{\overline a,\overline
  a'}=\begin{pmatrix}0&1\\1&0\end{pmatrix}\, .\label{r2}
\end{eqnarray}

\section{Vertex operators of chiral states}

In this section we present a derivation of the vertex operators
creating spectral flow images of chiral primary states. These
operators were proposed in \cite{Giribet:2007wp}.

\bigskip

\noindent{\it 3.1 NS sector}
\medskip

The Clebsch-Gordan
coefficients expanding the product representation ($\psi\Phi$)
 in (\ref{vertexfloww}) and (\ref{vfc}) are computed in the appendix. We find
\begin{equation}   (\psi^{|\hat h|}_{\hat m}\Phi^{h,w}_{H,M})^{}
_{{\cal H},{\cal M}}=
\sum_{\hat{m}=-\hat h}^{\hat h}
C^{{M},\hat{m},{\cal M}}_{H,\hat{h},{\cal H}}
\psi^{|\hat h|}_{\hat{m}}\Phi^{h,w}_{H,{M}}
\, ,\label{produsl1}
\end{equation}
where 
only the holomorphic part has been written and
\footnote{ We found convenient to denote the coefficients
$C^{{M},\hat{m},{\cal M}}_{H,\hat{h},{\cal H}}$ as
$\langle
H,M;\hat{h},\hat m|H,\hat{h};{\cal H},{\cal M}\rangle$ in the appendix.}
\begin{eqnarray}
C^{{M},\hat{m},{\cal M}}_{H,\hat{h},{\cal H}} &=&
\frac{({\cal M}+{\cal H})!}{(\hat{m}+|\hat{h}|)!({\cal
    M}-\hat{m}+H)!}\sum_{s=0}^
{\hat{m}+|\hat{h}|}
(-1)^{s-|\hat{h}|}\begin{pmatrix}\hat{m}+|\hat{h}|\\s\end{pmatrix}\frac{({\cal
      M}-s+
|\hat{h}|+H)!}{({\cal M}-s+{\cal H})!}\nonumber\\
&&\times \frac{(|\hat{h}|-H+{\cal M}-s-1)!}{({\cal M}-s-{\cal
      H}-1)!({\cal H}+|\hat{h}|-H)!}\, .
\label{coeffiseries1}
\end{eqnarray}

This can be rewritten using the generalized hypergeometric function
${}_3F_2(a,b,c;e,f|1)$ as
\begin{eqnarray}
C^{{M},\hat{m},{\cal M}}_{H,\hat{h},{\cal H}}&=&\frac{(-1)^{\hat{m}-|\hat{h}|}
\Gamma (-H+|\hat{h}|+{\cal M}) \Gamma
(H+|\hat{h}|+{\cal M}+1)}{\Gamma ({\cal H}-H+|\hat{h}|+1) \Gamma
({\cal M}-{\cal H})
   \Gamma (H+{\cal M}-\hat{m}+1) \Gamma (|\hat{h}|+\hat{m}+1)} 
\times\nonumber\\
   && _3F_2(-{\cal H}-{\cal M},{\cal H}-{\cal
  M}+1,-|\hat{h}|-\hat{m};-H-|\hat{h}|-
{\cal M},H-|\hat{h}|-{\cal M}+1;1)\, ,\nonumber
\end{eqnarray}
with the advantage
that it
can be represented
in terms of the Pochhammer double-loop contour integral,
possessing a unique analytic continuation in the complex
plane for all its indices \cite{Pochhammer, Holman}. Recall that the analogous
coefficients for SU(2) are related to these ones through analytic 
continuation.

For our purposes, it is convenient to
write the vertex operators in the $x-$basis, where the isospin can be
identified with the coordinates on the boundary.
This can be done using
\cite{Kutasov:1999}:
\begin{equation}
 { e^{-x J_0^-}\mathcal{O}(z)e^{x J_0^-} \equiv
\mathcal{O}(x,z).}\label{changbas}\end{equation}
Performing this operation  on the fermion fields, one gets in the unflowed
frame
\ber  e^{-x J_0^-}\psi^+(z)e^{x J_0^-}&=&\psi^+(x,z)\equiv \psi(x,z)\\
&=& {-2 x \psi^0(z)+\psi^+(z)+x^2 \psi^-(z),}
\eer
and in a generic $w$ frame
\begin{equation}
e^{-x J_0^-}\psi^{|\hat h|}_{\hat{m}=\hat h}(z)e^{x J_0^-}\equiv
\psi^{|\hat h|}(x,z)=
\sum_{\hat{m}=-\hat h}^{\hat h}
\frac{(-1)^{\hat{m}+\hat h}~\Gamma(2|\hat
  h|+1)}{\Gamma(\hat{m}+|\hat h|+1)
\Gamma(|\hat h|-\hat{m}+1)}\psi^{|\hat h|}
_{\hat{m}}x^{-\hat h+\hat m}\, .\label{psiw}
\end{equation}

Inserting
$H=m+\frac{k+2}{2}w$ and $\hat{h}=-w-1$ in (\ref{coeffiseries1})
we get
\begin{equation}
C^{{M},\hat{m},{\cal M}}_{H,\hat{h},{\cal
    H}}=(-1)^{\hat{m}+w+1}
\frac{\Gamma(2w+3)}
{\Gamma(\hat{m}+w+2)\Gamma(w-\hat{m}+2)}\, ,
\label{coef}
\end{equation}
 which coincide with the
coefficients in  (\ref{psiw}). Therefore,
 the SL(2,$\mathbb R)$ part of the chiral
vertex (\ref{vertexfloww}) may be written as
\begin{equation}
(\psi^{w+1}_{\hat m}\Phi^{h,w}_{H,M})^{}
_{{\cal H},{\cal M}}
=\sum_{\hat{m}=-w-1}^{w+1}(-1)^{\hat{m}+w+1}
\frac{\Gamma(2w+3)}
{\Gamma(\hat{m}+w+2)\Gamma(w-\hat{m}+2)}\psi^{w+1}_{\hat{m}}\Phi^{h,w}_{H,M}\, .
\label{expanssionphi}
\end{equation}
Expanding in modes, it is
easy to see that they may be expressed in the following factorized form
\begin{equation}
(\psi\Phi)^{h,w}_{{\cal H}}(x)\equiv\sum_{\cal M}
(\psi^{w+1}_{\hat m}\Phi^{h,w}_
{H,M})_{{\cal
      H},{\cal M}}\
x^{-{\cal H}+{\cal M}}=\psi^{w+1}(x)\Phi_{H}^{h,w}(x)\, .
\end{equation}
This factorization always occurs in  (\ref{vertexfloww}) when   $H$ and
$\hat{h}$ combine to produce a chiral state.

So far, we have restricted to the holomorphic SL(2,$\mathbb R)$
sector, but the same analysis applies   to
SU(2) \cite{Holman} and to their antiholomorphic parts.
Putting all  together, we get the following  vertex
operators creating spectral flow images of chiral primary
 states in arbitrary spectral flow frames
\ber
\mathcal{W}^{h,w}_{{\cal H}\overline {\cal H}}(x,y,\overline x,\overline y)
&=&
e^{-\varphi}\ \Phi^{h,w}_{{ H}\overline{H}}(x,\overline x)
\psi^{w+1}(x) \overline\psi^{w+1}(\overline x)
V^{h-1,w}_{{ J}\overline {J}}(y,\overline y)\chi^w(y)\overline
\chi^w(\overline y),
\label{wvof}\\
{\cal Y}^{h, w}_{{\cal H}\overline {\cal
    H}}(x, y,\overline x, \overline y)&=&
e^{-\varphi}\Phi^{h,w}_{H\overline H }(x,\overline
    x)\psi^w(x)\overline \psi^w(\overline x)
\chi^{w+1}(y)\overline \chi^{w+1}(\overline y)
 V^{h-1,w}_{J,\overline
J}(y, \overline y)
\,
,\label{chivof}
\eer
with $J=H-2w, \overline J=\overline H-2w$, ${\cal H}=-J-1,
    \overline{\cal H}=-\overline J-1$.

\medskip
\noindent{\it 3.2 Ramond sector}

\medskip
The product representation needed to construct the vertex operators (\ref{vertexram})
in the Ramond sector
can be expanded as
\ber
(S^{\hat j}_{\hat m,\hat m'}
\Phi^{h,w}_{H,M} V^{j,w}_{J,M'})_{{\cal H}, {\cal M}, {\cal J}, {\cal M}'}&=&
\sum_{\hat{m},\hat{m'}=-\hat h}^{\hat h}(S^{\hat j}_{\hat m,\hat m'}
\Phi^{h,w}_{H,M} V^{j,w}_{J,M'})
C^{({M},\hat{m},{\cal M}),({M'},\hat{m'},{\cal M'})}_{(H,\hat{h},{\cal
    H}),
(J,\hat{h},{\cal J})}~~~~~~~~~~~~~~~~~\nonumber\\
&\equiv&\sum_{\hat{m}=-\hat h}^{\hat h}
(S^{\hat j}_{\hat m}
\Phi^{h,w}_{H,M})
C^{{M},\hat{m},{\cal M}}_{H,\hat{h},{\cal H}}\otimes\sum_{\hat{m'}=
-\hat h}^{\hat h}(S^{\hat j}_{\hat m'}V^{j,w}_{J,M'})
C^{{M'},\hat{m'},{\cal M'}}_{J,\hat{h},{\cal J}}
,\label{produsspf}
\eer
$i.e.$ the SL(2) and SU(2) parts factorize. The
Clebsch-Gordan coefficients $C^{{M},\hat{m},{\cal M}}_{H,\hat{h},{\cal
    H}}$
can be computed from (\ref{coeffiseries1}) taking
$H=m+\frac{k+2}{2}w$ and $\hat{h}=-w-\frac 12$. Using (\ref{changbas}), it
is easy to see that the triple product factorizes
in the $x-$basis as
\ber
 (S\Phi V)^{h,w}_{{\cal H},{\cal J}}(x,y)&\equiv &\sum_{{\cal M},{\cal
    M'}}
(S^{w+\frac 12}_{\hat m,\hat m'}\Phi^{h,w}_
{H,M}V^{j,w}_{J;M'})^{}_{{\cal
      H},{\cal M},{\cal J},{\cal M}'}\
x^{-{\cal H}+{\cal M}}y^{{\cal J}+{\cal M'}}\nonumber\\
&=&S^{w+\frac 12}(x,y)\Phi_{H}^{h,w}(x)V_{J}^{j,w}(y)\, ,
\eer
where \footnote{These spin fields are denoted $S^-_w(x,y)$ in 
\cite{Giribet:2007wp}}

\ber
S^{w+\frac 12}(x,y)&\equiv&
\sum_{\hat{m},\hat{m'}=-(w+\frac 12)}^{w+\frac 12}
\left[\frac{(-1)^{\hat{m}+w+\frac 12}~\Gamma(2w+2)}
{\Gamma(\hat{m}+w+\frac 32)
\Gamma(w
-\hat{m}+\frac 32)}\frac{(-1)^{\hat{m}'+w+\frac 12}~\Gamma(2w+2)}
{\Gamma(\hat{m}'+w+\frac 32)
\Gamma(w
-\hat{m}'+\frac 32)}\right]\nonumber\\
&&~~~~~~~~~~~~~~~~~~~~~ \times ~S^{w+\frac 12}_{\hat{m},\hat{m}'}\,
x^{\hat{m}+w+\frac 12}y^{\hat{m}'+w+\frac 12}\, .\label{Sw}
\eer
Taking into account the antiholomorphic part, the vertex operators
creating spectral flow images of chiral primary states in the
Ramond sector are given by

\begin{equation}\label{vertex.ramonx}
{\cal R}^{\pm,h,w}_{\cal H\overline{\cal H}}(x,\overline x,
y,\overline y)=
e^{-\frac{\varphi} 2}
S^{w+\frac 12}(x, y)\overline S^{w+\frac 12}(\overline x, \overline y)
\Phi^{h,w}_{H,\overline{H}}(x,\overline x) V^{j,w}_{J,\overline
  J}(y,\overline y)
e^{\pm \frac i2(\hat H_4-\hat H_5)}e^{\pm \frac i2(\hat{\overline
    H}_4-
\hat{\overline H}_5)}\, .
\end{equation}

The expressions (\ref{wvof}), (\ref{chivof}) and (\ref{vertex.ramonx})
that we deduced here appeared previously  in
\cite{Giribet:2007wp}.

\section{Three-point functions of chiral states}

In this section we compute $w-$conserving three-point functions involving
spectral flow images of chiral primary states. We restrict to the so called
extremal correlators, satisfying  $j_n=j_m+j_l$.
\medskip

\noindent{\it 4.1 NS-NS-NS three-point functions }

Let us start by evaluating the following amplitudes
\begin{equation}  \label{3pcf1}
{\cal A}_{3}=g_s^{-2}\left<\mathcal{W}^{h_1,w_1}_{{\cal H}_1, \overline 
{\cal H}_1}
(x_1,y_1,\overline{x}_1,\overline y_1)
\mathcal{W}^{h_2,w_2}_{{\cal H}_2,\overline {\cal H}_2}(x_2,y_2,\overline{x}_2,
\overline y_2)
\mathcal{W}_{{\cal H}_3,\overline {\cal H}_3}^{h_3,w_3}
(x_3,y_3,\overline{x}_3,\overline y_3)\right>_{S^2}\, ,
\end{equation}
\begin{equation}\label{3pcf2}
{\cal A'}_{3}=g_s^{-2}\left<\mathcal{Y}^{h_1,w_1}_{{\cal
    H}_1,\overline {\cal H}_1}(x_1,y_1,\overline
{x}_1,\overline y_1)\mathcal{W}^{h_2,w_2}_{{\cal
    H}_2\overline {\cal H}_2}(x_2,y_2,\overline{x}_2,\overline y_2)
\mathcal{Y}^{h_3,w_3}_{{\cal H}_3,\overline {\cal H}_3}
(x_3,y_3,\overline{x}_3,\overline y_3)\right>_{S^2}\, .
\end{equation}

The vertices ${\cal W}^{h,w}_{{\cal H},\overline{\cal H}}$,
${\cal Y}^{h,w}_{{\cal H},\overline{\cal H}}$
were defined in
the $-1$ ghost picture. To have
 total ghost number  $-2$, as required on the sphere,
we change the picture of an unflowed operator for simplicity, $i.e.$
\cite{Gaberdiel, Pakman}
\begin{equation}
 \mathcal{W}^{(0)}_{h}(x, y, \overline x, \overline y) =\left [\left( (1-
h)\hat{\jmath}(x) + j(x) + \frac{2}{k} \psi(x) \chi_a(y) P^a_y
\right)\times c.c.\right ]
\Phi_{h}(x,\overline x) V_{h-1}(y,\overline y)\, , \label{zppo}
\end{equation}
\begin{equation}
 \mathcal{Y}^{(0)}_{h}(x, y, \overline x, \overline y) =\left [\left(
h\hat{k}(y) + k(y) + \frac{2}{k} \chi(y) \psi_a(x) D^a_x
\right)\times c.c.\right ]
\Phi_{h}(x,\overline x) V_{h-1}(y,\overline y)\, . \label{yppo} \end{equation}
As discussed in detail below, this restriction is not strictly 
necessary to evaluate
(\ref{3pcf2}), but further knowledge on
 spectral flowed affine 
representations than is currently available is needed  to
compute (\ref{3pcf1}) in a more general situation. In any case, we
shall see that
including an unflowed operator
does not imply any loss of generality for correlators
involving spectral flow images of chiral primary states in the SL(2, $\mathbb
R$) sector.

Replacing (\ref{zppo}) in
(\ref{3pcf1}), ${\cal A}_3$ explicitly reads
 \ber\label{a3} {\cal A}_{3}&=& g_s^{-2}\left\langle e^{-\varphi(z_1,\bar
   z_1)}e^{-\varphi(z_2,\bar z_2  )}\right\rangle
\left\langle
 V^{h_1-1,w}_{J_1,\overline J_1}
(y_1,\overline y_1)V^{h_2-1,w}_{J_2, \overline J_2}(y_2,\overline y_2)
V_{h_3-1}(y_3,\overline y_3)\right\rangle
\nonumber\\
&\times &
\left\langle\overline \chi^{w}(\overline
y_1)\overline
\chi^{w}(\overline y_2)\right\rangle\left\langle
\chi^{w}(y_1)\chi^{w}(y_2)\right\rangle
\left\langle\Phi^{h_1,w}_{H_1,\overline H_1}(x_1,\overline
x_1)\psi^{w+1}(x_1)\overline\psi^{w+1}(\overline x_1)
\Phi^{h_2,w}_{H_2,\overline H_2}(x_2,\overline
x_2)\right.\nonumber \\
&\times &\left.\psi^{w+1}(x_2)\overline\psi^{w+1}(\overline
x_2)\left \{(1-h_3)\hat{\jmath}(x_3)+j(x_3)\right \}\left \{(1-h_3)
\hat{\jmath}(\overline x_3)+j(\overline x_3)\right \}\Phi_{h_3}(x_3,\overline
x_3)\right\rangle\nonumber\, ,
\eer
and inserting (\ref{yppo}) into (\ref{3pcf2}) and using 
$\psi_a(x) D^a_x=\frac 12(\psi(x)\partial_x+h\partial_x\psi(x))$ we get
\ber  {\cal A'}_{3}&=& g_s^{-2}
\left\langle e^{-\varphi(z_1,\bar z_1)}e^{-\varphi(z_2,\bar z_2)}\right\rangle
\left\langle V^{h_1-1,w}_{J_1,\overline J_1}(y_1,\overline y_1)
V^{h_2-1,w}_{J_2,\overline J_2}(y_2,\overline
y_2)
V_{h_3-1}(y_3,\overline y_3)\right\rangle\nonumber \\
&\times &\left\{
\left [ \left\langle\psi^{w}(x_1)\psi^{w+1}(x_2)\psi(x_3)\right\rangle
\partial_{x_3}
\left\langle\Phi^{w}_{{ H}_1,\overline { H}_1}(x_1,\overline x_1)
\Phi^{w}_{{ H}_2,\overline {H}_2}(x_2,\overline x_2)
\Phi_{h_3}(x_3,\overline x_3)\right\rangle
 +
\right.\right.
\nonumber\\
&&~~~h_3\left.\left.\left\langle\psi^{w}(x_1)\psi^{w+1}(x_2)
\partial_{x_3}
\psi(x_3)
\right\rangle\left\langle\Phi^{w}_{H_1,\overline H_1}(x_1,\overline x_1)
\Phi^{w}_{{H}_2,\overline{H}_2}(x_2,\overline x_2)\Phi_{h_3}(x_3,\overline
x_3)\right\rangle\right ]\right.\nonumber\\
&\times&\left.
\left\langle\chi^{w+1}(y_1)\chi^{w}(y_2)\chi(y_3)\right\rangle
\times c.c.\right\}
\label{c'3}\, , \eer
where $w=w_1=w_2$.

The
SU(2)
and fermionic
expectation values were discussed in \cite{Giribet:2007wp}.
We now compute the SL(2,$\mathbb R)$ correlators,
applying the technique developed in \cite{mo3}.

From the integral transform
\begin{equation}
\Phi_{H,M,\overline H,\overline M}^{h,w}=\int{d^2x}x^{H-M-1}\overline
x^{\overline H-\overline M-1}\Phi_{H,\overline H}^{h,w}(x,\overline x)\;
,\label{itf}
\end{equation}
a generic three-point function in the $x-$basis, $e.g.$ (we
omit the $z$
dependence for short)
\begin{eqnarray}
\left\langle\Phi^{h_1,w_1}_{H_1,\overline H_1}(x_1,\overline x_1)
\Phi^{h_2,w_2}_{H_2,\overline H_2}(x_2,\overline x_2)
\Phi^{h_3,w_3}_{H_3,\overline
H_3} (x_3,\overline x_3)\right\rangle
=
D(H_i,\overline
H_i)\left (x_{12}^{-H_{12}} x_{23}^{-H_{23}}
x_{13}^{-H_{13}}\times c.c.\right )
\; ,\label{418}
\end{eqnarray}
($c.c.$ stands for the antiholomorphic
dependence),
can be transformed to the $m-$basis as
\begin{equation}
\left\langle\prod_{i=1}^3\Phi_{H_i,M_i,\overline H_i,\overline
  M_i}^{h_i,w_i'}
\right
\rangle=(2\pi)^2
D(H_i,\overline H_i)W(H_i,M_i,\overline H_i,\overline M_i)
\delta^2(M_1+M_2+M_3),
\end{equation}
where
\begin{eqnarray}
W(H_i,M_i,\overline H_i,\overline M_i)&=&\int d^2x_1 d^2x_2 x_1^{H_1-M_1-1}
x_2^{H_2-M_2-1}\overline x_1^{H_1-\overline M_1-1} \overline
x_2^{H_2-\overline M_2-1}|x_{12}|^{-2H_{12}}\nonumber\\
&&~~~~~~~~~~~~~~\times ~|1-x_1|^{-2H_{13}}|1-x_2|^{-2H_{23}}\;
.\label{ww}
\end{eqnarray}

Recall that the spectral flow with $w>0$ ($w<0$) turns  primary states
of the current algebra into  lowest- (highest-) weight states of a global
representation with $H=M=m+\frac {k+2}2 w$, $\overline H=\overline
M=\overline
m+\frac {k+2}2 w$
($H=-M=-m+\frac {k+2}2|w|$, $\overline H=-\overline M=-\overline
m+\frac {k+2}2|w|$) \footnote{The spectral flow labels $w$ and $w'$ 
for highest/lowest weight states of global representations in the 
$x-$ and $m-$basis, respectively,  may be related as $w'=\frac MHw$.}.
 Therefore,
we are interested in the
residue of the poles at say, $H_1=M_1, \overline H_1=\overline M_1$
and $H_2=-M_2, \overline H_2=-\overline M_2$.
This is obtained by taking the $x_1,\overline x_1\rightarrow 0$ and $x_2, 
\overline
x_2\rightarrow \infty$ limits in the integrand of $W(H_i,\overline H_i,M_i,
\overline
M_i)$,
 which  simply gives
\begin{eqnarray}
\left\langle\prod_{i=1}^3\Phi_{H_i,M_i,\overline H_i,\overline M_i}^{h_i,w_i'}
\right
\rangle = (2\pi)^2
V^2_{conf}\delta^2(M_1+M_2+M_3)D(H_i,\overline H_i),~~~
\label{3ptm}
\end{eqnarray}
where $V_{conf}=\int dx^2/|x|^2$.

On the other hand, it is well known that
 spectral flow
 preserving $n-$point functions in the $m-$basis
are related to correlators involving
 only unflowed operators as
\begin{equation}
\left\langle\prod_{i=1}^n\Phi_{H_i,M_i,\overline
   H_i,\overline M_i}^{h_i,w_i}(z_i,\overline
  z_i)\right\rangle_{\sum_iw_i=0}=\prod_{j<i}(z_{ij})^{-w_jm_i-w_im_j-
\frac {k+2}2 w_iw_j}\times c.c.
  \left\langle\prod_{i=1}^n\Phi_{h_i,m_i,\overline m_i}^{w_i=0}(z_i,\overline
  z_i)\right\rangle\; ,\label{ribault}
\end{equation}
and three-point functions of $w=0$ primary states
 have the following form \cite{tesch1, satoh}:
\begin{eqnarray}
\left\langle\prod_{i=1}^3\Phi_{h_i,m_i,\overline
  m_i}^{w_i=0}(z_i,\overline z_i)\right\rangle = (2\pi)^2
  \delta^2(\sum_im_i)W(h_i,m_i,\overline m_i)C(h_i)
|z_{12}|^{-2\Delta_{12}}|z_{13}|^{-2\Delta_{13}}|z_{23}|^{-2\Delta_{23}}
  ,\nonumber\\
\label{3pfm}
\end{eqnarray}
with
\begin{eqnarray}
 C(h_1,h_2,h_3)=-\frac{G(1-h_1-h_2-h_3)G(-h_{12})G(-h_{13})G(-h_{23})}{2\pi^2
\nu^{h_1+h_2+h_3-1}\gamma\left (\frac{k+1}{k}\right
)G(-1)G(1-2h_1)G(1-2h_2)G(1-2h_3)}\; ,
\end{eqnarray}
where
 $G(h)=k^{\frac{j(k+1-h)}{2k}}\Gamma_2(-h|1,k)\Gamma_2(k+1+h|1,k)$,
 $\Gamma_2$ being the Barnes double gamma function and
$\Delta_{12}=\Delta_1+\Delta_2-\Delta_3$, $h_{12}=h_1+h_2-h_3$,
 etc.


Comparing with
(\ref{3ptm}), one finds that the
three-point
functions involving spectral flow images of primary
 operators in arbitrary $w-$sectors in the
$x-$basis corresponding to $w-$preserving amplitudes in the $m-$basis
 are given by
\footnote{This correlation function was directly computed  in the
  $x-$basis
in \cite{hmn}
in the particular
case $w_1=w_2=1$, $w_3=0$ using the definition of $w=1$ vertex 
operators given in
  \cite{mo3}.  Here we have used
a different technique which is useful to evaluate correlators
  involving fields in arbitrary $w$ sectors and, specially,
  expectation values including currents.}
\begin{eqnarray}
&&\left\langle\Phi_{H_1, \overline H_1}^{h_1,w_1} (x_1,\overline
x_1) \Phi_{H_2, \overline H_2}^{h_2,w_2} (x_2,\overline
x_2)\Phi_{H_3,\overline H_3}^{h_3,w_3} (x_3,\overline
x_3)\right\rangle
\nonumber\\
&&~~~~~~~~ ~~~~~~~
=~\frac 1{V_{conf}^2}W(h_i,m_i,\overline m_i)
C(h_i)
x_{12}^{-H_{12}} x_{13}^{-H_{13}}x_{23}^{-H_{23}}
\overline x_{12}^{-\overline H_{12}} \overline
x_{13}^{-\overline H_{13}}\overline x_{23}^{-\overline H_{23}}
\;.~~~~~~~~~~~
\label{3pfx}
\end{eqnarray}
Recall that this result holds for operators satisfying $m_1+m_2+m_3=0$.

As discussed above,
for highest/lowest weight states the function
$W(h_i,m_i,\overline m_i)$ develops poles
which cancel the factor $V_{conf}^{-2}$.
Taking, for instance,
 a chiral field at $x_1, \overline x_1$  and an antichiral one at
$x_2, \overline x_2$, $i.e.$ $m_1=\overline
m_1=h_1,
m_2=\overline m_2=-h_2$, the residue
of the double pole is just one, and we obtain \footnote{Normalizing 
the two-point functions of these operators to the identity,
this result agrees
with the prediction formulated in \cite{Giribet:2007wp} when the correlator
involves one unflowed state. Three flowed chiral primary
operators obeying $m_1+m_2+m_3=0$ cannot meet the condition
$h_3=h_1+h_2-1$ under which the prediction of \cite{Giribet:2007wp} holds.}
\ber
{\cal A}^1_3\equiv\<\Phi^{h_1,w_1}_{H_1,\overline
H_1}(x_1,\overline x_1)\Phi^{h_2,w_2}_{H_2,\overline H_2}(x_2,\overline
x_2)\Phi^{h_3,w_3}_{H_3,\overline H_3}(x_3,\overline x_3)\>=C(h_i)
x_{12}^{-H_{12}} x_{13}^{-H_{13}}x_{23}^{-H_{23}}
\overline x_{12}^{-\overline H_{12}} \overline
x_{13}^{-\overline H_{13}}\overline x_{23}^{-\overline H_{23}}
.\nonumber
\eer

The following expectation value is also needed to evaluate ${\cal A}_3$:
\ber
{\cal A}^2_3&\equiv&\<\Phi^{h_1,w_1}_{H_1,\overline H_1}(x_1,\overline
x_1)\Phi^{h_2,w_2}_{H_2,\overline H_2}(x_2,\overline x_2)j(
x_3)\Phi_{h_3}(x_3,\overline
x_3)\>\, .\nonumber
\eer

The OPE $j(x)\Phi_{H,\overline H}^{h,w}(x',\overline
x')$  is only known so far for $w=1$ fields
\cite{mo3}, namely
\ber\label{opeflow} j(x',z')\Phi_{H,\overline H}^{h,w=1}(x,\overline
x;z,\overline z)&=&
(m-h+1)\frac{(x-x')^2}{(z-z')^2}\Phi^{h,w=1}_{H+1,\overline
  H}(x,\overline x;z,\overline z)~~~~~~~~~~~~~~~\nonumber\\
&& +~ \frac{1}{z'-z}
\left[2H(x-x')+(x-x')^2\partial_x\right]\Phi^{h,w=1}_{H,\overline
  H}(x,\overline x;z,\overline z)\, .~~~\label{jphi}
\eer
Therefore, we restrict to this case.
Inserting
(\ref{jphi}) into ${\cal A}_3^2$, one gets
 \ber\label{w13pf}
{\cal A}_3^2&=& (1-h_1+m_1)
\frac{(x_1-x_3)^2}{(z_1-z_3)^2}
<\Phi^{h_1,w=1}_{H_1+1,\overline
H_1}(x_1,\bar x_1)
\Phi^{h_2,w=1}_{H_2,\overline H_2}(x_2,\bar x_2)\Phi_{h_3}(x_3,\bar x_3)>\nonumber\\
&&
+~(1-h_2-m_2)\frac{(x_2-x_3)^2}{(z_2-z_3)^2}<\Phi^{h_1,w=1}_{H_1,\overline
H_1}(x_1,\bar x_1)
\Phi^{w=1,h_2}_{H_2+1,\overline H_2}(x_2,\bar x_2)\Phi_{h_3}(x_3,\bar x_3)>\nonumber\\
&&+~\frac{1}{z_3-z_1}\left[2H_1(x_1-x_3)+(x_1-x_3)^2\partial_{x_1}\right]
{\cal A}_3^1
\nonumber\\
&&+~\frac{1}{z_3-z_2}\left[2H_2(x_2-x_3)+(x_2-x_3)^2\partial_{x_2}\right]
{\cal A}_3^1\, .
\nonumber \eer


The first two terms are easily evaluated using
the  procedure discussed above and we get
\begin{eqnarray}
\left\langle\Phi^{h_1,w=1}_{H_1+1,\overline
H_1}(x_1,\bar x_1)
\Phi^{h_2,w=1}_{H_2,\overline H_2}(x_2,\bar x_2)\Phi_{h_3}(x_3,\bar
x_3)\right\rangle=
~W(h_i, m_1=h_1+1, m_2=-h_2,m_3)
\nonumber\\
\times~ V_{conf}^{-2}~C(h_i)~
x_{12}^{-H_{12}-1} x_{13}^{-H_{13}-1}x_{23}^{-H_{23}+1}
\overline x_{12}^{-\overline H_{12}} \overline
x_{13}^{-\overline H_{13}}\overline x_{23}^{-\overline H_{23}}\, ,
\end{eqnarray}
where
\begin{equation}
W(h_i, m_1=h_1+1, m_2=-h_2,m_3)= V_{conf}^2
\frac{h_{13}}{m_1-h_1+1}\, ,
\end{equation}
and similarly,
\begin{eqnarray}
&&\left\langle\Phi^{h_1,w=1}_{H_1,\overline
H_1}(x_1,\overline x_1)
\Phi^{h_2,w=1}_{H_2+1,\overline H_2}(x_2,\overline x_2)\Phi_{h_3}(x_3,\overline
x_3)\right\rangle=~C(h_i)\frac{h_{23}}{1-h_2-m_2}~~~~~~~~~
\nonumber\\
&&~~~~~~~~~~~~~~~~~~~~~~~~~~~~~~~~~~~~~~~\times~
x_{12}^{-H_{12}-1} x_{13}^{-H_{13}+1}x_{23}^{-H_{23}-1}
\overline x_{12}^{-\overline H_{12}} \overline
x_{13}^{-\overline H_{13}}\overline x_{23}^{-\overline H_{23}}\, .
\end{eqnarray}
Putting all together, we obtain
\begin{eqnarray}
{\cal A}_3^2 ~= ~(3h_3-H_1-H_2)C(h_i)~x_{12}^{-H_{12}-1} x_{13}^{-H_{13}+1}
x_{23}^{-H_{23}+1}
\overline x_{12}^{-\overline H_{12}} \overline
x_{13}^{-\overline H_{13}}\overline x_{23}^{-\overline
  H_{23}}\nonumber\\
\times~z_{12}^{-\Delta_{12}+1}
z_{13}^{-\Delta_{13}-1}z_{23}^{-\Delta_{23}-1}
\overline z_{12}^{-\overline\Delta_{12}}
z_{13}^{-\overline\Delta_{13}}\overline
z_{23}^{-\overline\Delta_{23}}\, ,
\end{eqnarray}
and analogously for the term containing the antiholomorphic current
$\overline j(x)$ in ${\cal A}_3$.

To write down the final result, let us recall the fermionic and SU(2)
correlators
(see \cite{Giribet:2007wp} for details).
\begin{equation}\label{2pfflow} <\psi^{w+1}(x_1)\psi^{w+1}(x_2)>=\frac k2
\frac{x_{12}^{2(w+1)}}{z_{12}^{(w+1)^2}}~~,
 \end{equation}
\begin{eqnarray}\label{3pcflow} <\psi^{w+1}(x_1)\psi^{w+1}(x_2)\hat{\jmath}
(x_3)>
&=&\sum_{i=1}^2\frac{1}{z_{3i}}\left[2(w+1)x_{3i}+(x_{3i})^2
\partial_{x_i}\right]<\psi^{w+1}(x_1)\psi^{w+1}(x_2)>\nonumber\\
&=&
k(w+1)\frac{x_{13}x_{23}}
{x_{12}}\frac{z_{12}}{z_{13}z_{23}}\frac{x_{12}^{2(w+1)
    }}{z_{12}^{(w+1)^2}}\, ,
\end{eqnarray}
\begin{eqnarray}
  <\psi^{w}(x_1)\psi^{w+1}(x_2)\psi(x_3)>
=k\frac{x_{12}^{2w}x_{23}^{2}z_{13}^w}{z_{12}^{w^2+w}z_{23}^{w+1}}\, .
\end{eqnarray}

Similar expressions are obtained for $\chi^w$.

In the
  SU(2) WZNW model, normalizing the two$-$point functions 
as
\begin{equation} \langle V_{j_1}(y_1,\overline y_1;z_1,\overline
z_1)V_{j_2}(y_2,\overline y_2;z_2,\overline
z_2)=\delta_{j_1j_2}\frac{|y_{12}|^{2j_1}}{|z_{12}|^{4\Delta_{j_1}}}\,
,\end{equation}
the three$-$point functions 
are given by \cite{Zamolod}
\begin{equation}\label{3ptphiprime} \langle 
V_{j_1}(y_1, \bar y_1; z_1,\bar z_1)
V_{j_2}(y_2, \bar y_2;z_2, \bar z_2) V_{j_3}(y_3 ,\bar y_3;z_3,
\bar z_3) \rangle = C'(j_1, j_2, j_3) \, \prod_{i<j}
\frac{|y_{ij}|^{2j_{ij}}} {|z_{ij}|^{2\Delta_{ij}}} \ ,
\end{equation} for $j_n\le j_m+j_l$, where
\begin{equation} \label{CSU2} C'(j_1,j_2,j_3) =
\sqrt{\frac{\gamma({\textstyle\frac{1}{k}})}{
\gamma(\frac{2j_1+1}{k})\gamma(\frac{2j_2+1}{k})
\gamma(\frac{2j_3+1}{k})}}  \, \frac{P(j_1+j_2+j_3+1)\,
P(j_{12})\, P(j_{23})\, P(j_{31})}{ P(2j_1)\, P(2j_2)\, P(2j_3)}\nonumber
\end{equation} with $ P(j)=\prod_{m=1}^{j}
\gamma({\textstyle\frac{m}{k}}), P(0)=1$.

 As argued in \cite{Giribet:2007wp},
 the structure constants 
for spectral 
flowed chiral fields in SU(2)  are also given by $C'(j_i)$ for $j_n=j_m+j_l$.
Therefore, collecting all the contributions and suppressing the $x-$ and
$z-$dependence for short, we get
\begin{equation}\label{final3pf}
{\cal A}_3=g_s^{-2}\frac {k^2}4
\left |\mathcal{H}_1+
\mathcal{H}_2+\mathcal{H}_3+1\right |^2C'(j_i)C(h_i)\,
, \end{equation}
\ber\label{final3pff}
{\cal A'}_3=g_s^{-2}\frac {k^2}4
\left |\mathcal{H}_1-\mathcal{H}_2+\mathcal{H}_3-1\right |^2C'(j_i)
C(h_i)\,.\eer
As shown in \cite{Gaberdiel, Pakman},
\begin{equation}
C'(j_i)C(h_i)=\sqrt{B(h_1)B(h_2)B(h_3)}, \qquad B(h_i)=\frac
{k}{4\pi^3}
\frac{\nu^{1-2h_i}}{\gamma\left (\frac{2h_i-1}{k}\right )}\, ,\quad
\nu=\pi\frac{\Gamma\left (1-\frac 1{k}\right )}{\Gamma\left (1+\frac
  1{k}\right )}\, ,
\end{equation}
with $\gamma(x)=\frac{\Gamma(x)}{\Gamma(1-x)}$.

In order to compare these results with the conjectured dual counterparts,
the two-point functions must be normalized to the identity.
Taking into account that in the SL(2,$\mathbb R)$ sector they are given by
\cite{mo3} \ber
 \langle\Phi_{{H},\overline {H}}^{h,w}(x_1, \overline
x_1) \Phi_{ {H},\overline {H}}^{h,w} (x_2,
\overline x_2)\rangle =
 g_s^{-2} (2h-1+k w)B(h)
x_{12}^{-2{H}}\overline x_{12}^{-2\overline {H}}\, , \eer
 the normalized chiral operators are defined as:
\begin{equation}\label{normalized}
\mathbb{W}^{h,w}_{{\cal H},\overline {\cal H}}(x,\bar{x})
\equiv\frac{4g_s\mathcal{W}^{h,w}_{{\cal H},\overline {\cal
H}}(x,\bar{x})} {k^2\sqrt{B(h)(2h-1+kw)}} ,\qquad
\mathbb{Y}^{h,w}_{{\cal H}\overline {\cal H}}(x,\bar{x})\equiv
\frac{4g_s\mathcal{Y}^{h,w}_{{\cal
      H}\overline {\cal H}}(x,\bar{x})}
{k^2\sqrt{B(h)(2h-1+kw)}}\, .
\end{equation}

 Omitting the standard dependence on the coordinates, we thus get
\ber\langle\mathbb{W}^{h_1,w}_{{\cal H}_1\overline {\cal
    H}_1}\mathbb{W}^{h_2,w}_{{\cal H}_2\overline {\cal
    H}_2}\mathbb{W}^{(0)}_{h_3}
\rangle&=&\frac{4g_s}{k^2}
\frac{\left |\mathcal{H}_1+\mathcal{H}_2+\mathcal{H}_3+1\right |^2}
{\sqrt{(2h_1-1+kw)(2h_2-1+kw)(2h_3-1)}}\, ,
\label{final3pfw}\eer
\ber
\langle\mathbb{Y}^{h_1,w}_{{\cal H}_1\overline{\cal
    H}_1}\mathbb{W}^{h_2,w}_{{\cal H}_2\overline {\cal H}_2}
\mathbb{Y}^{(0)}_{h_3}\rangle&=& \frac{4g_s}{k^2}
\frac{\left |\mathcal{H}_1-\mathcal{H}_2+\mathcal{H}_3-1\right |^2}
{\sqrt{(2h_1-1+kw)(2h_2-1+kw)(2h_3-1)}}\, .
\label{final3pfx}\eer

While (\ref{final3pfw}) was obtained for $w=1$, (\ref{final3pfx})
holds for arbitrary $w$.

These three-point functions involve one unflowed operator. We
restricted to this case for simplicity. However,
notice that
when the three operators are  spectral flow images of chiral
primaries of SL(2,$\mathbb R)$ or the unflowed operator creates a
highest/lowest weight primary state, the  condition $h_i=\pm m_i$ together with
the requirement $m_1+m_2+m_3=0$ imply, for example,
$h_2=h_1+h_3$. Combined with the chirality condition $j_i=h_i-1$, this
gives $j_2=j_1+j_3+1$ which violates the triangular inequality
$j_2\le j_1+j_3$ of the SU(2) WZNW model. Therefore the SU(2)
factor gives a zero for the whole three-point function. 
This conclusion does not  apply when  the unflowed operator obeys
$h_3\ne\pm m_3$. Therefore, the results
(\ref{final3pfw}) 
and (\ref{final3pfx}) hold for amplitudes containing
two flowed and one unflowed chiral primary 
operators as long as the latter does not create a
highest/lowest weight state in the SL(2, $\mathbb R)$ sector.

Let us now compare these results with the  correlators
in the dual
theory. The level $k$ is identified with N$_5$ \cite{ms} and
$g_s^2=\frac {{\rm N}_5}{{\rm N}_1}Vol({\rm T}^4)$ \cite{Kutasov98, 
Kutasov:1999, kll}, so
these correlation functions scale as N$^{-1/2}$ 
in the large N limit. 
Recall
that the  chiral string states
${\cal W}^{h,w}_{{\cal H}\overline {\cal H}}$,
${\cal Y}^{h,w}_{{\cal H}\overline {\cal H}}$ have been identified with
the chiral operators ${\cal O}^{--}_{n,\overline n}$, 
${\cal O}^{++}_{n,\overline n}$ 
of the SCFT, 
respectively
\cite{Gaberdiel,
Pakman, Giribet:2007wp}. 
Moreover, the proposed identification
between the quantum numbers of $\mathbb{W}^{h,w}_{{\cal H}\overline {\cal H}}$
and those of ${\cal O}^{--}_n(x,\bar x)$ is the following
\cite{Gaberdiel,
Pakman, Giribet:2007wp}
\begin{equation}
{\cal H}_n=\frac{n-1}{2}=h-1+\frac{k}{2}w\qquad \Rightarrow\qquad
n=2h-1+k w\, ,\end{equation} and for
$\mathbb{Y}^{h,w}_{{\cal H}\overline {\cal H}}$ and ${\cal O}^{++}_n(x,\bar
x)$ it is
\begin{equation}
{\cal H}_n=\frac{n+1}{2}=h+\frac{k}{2}w\qquad \Rightarrow\qquad
n=2h-1+k w\, .\end{equation} Replacing these values of $n$ in the
boundary three-point functions (\ref{3pfPS}), one gets at leading order
\ber\langle
{\cal O}^{--}_{n_1}{\cal O}^{--}_{n_2} {\cal O}^{--\dagger}_{n_3}\rangle&=&
\frac 1{\sqrt {\rm N}}
\frac{\left|{\cal H}_1+{\cal H}_2+{\cal H}_3+1\right|^2}
{\sqrt{(2h_1-1+kw_1)(2h_2-1+kw_2)(2h_3-1)}}\, ,\label{1c}\\
\langle {\cal O}^{++}_{n_1}{\cal O}^{--}_{n_2}{\cal O}^{++\dagger}_{n_3}
\rangle&=&\frac 1{\sqrt{\rm N}}
\frac{\left|{\cal H}_1-{\cal H}_2+{\cal H}_3-1\right|^2}
{\sqrt{(2h_1-1+kw_1)(2h_2-1+kw_2)(2h_3-1)}}\,\label{2c}
,\eer
in perfect agreement with
 (\ref{final3pfw}) and (\ref{final3pfx}), respectively. Furthermore, using the
bulk-to-boundary dictionary, one can verify that the boundary
correlators corresponding to three spectral flow images of chiral
 primary operators is zero
because in both cases (\ref{1c}) and (\ref{2c}) 
the relation $h_2=h_1+h_3$ implies $n_2=n_1+n_3$, which
violates  the U(1) charge by one unit.

Two other correlators can be considered in the string theory
corresponding to the vanishing correlators 
$(\epsilon_{n_1},\epsilon_{n_2},\epsilon_{n_3})=(+,+,+)$
and $(-,-,+)$ in the boundary CFT, namely
$\langle\prod_{i=1}^3 {\cal Y}^{h_i,w_i}_{{\cal H}_i,\overline 
{\cal H}_i}\rangle$ 
and $\langle\prod_{i=1}^2 {\cal W}^{h_i,w_i}_{{\cal H}_i,\overline 
{\cal H}_i}
{\cal Y}^{h_3,w_3}_{{\cal H}_3,\overline 
{\cal H}_3}
\rangle$. It is easy to see that they 
violate the SU(2) charge conservation in the case $j_2=j_1+j_3$ that we are considering,
and therefore they also vanish.

\bigskip

\noindent{\it 4.2 R-R-NS three-point functions}

\medskip

The chiral states ${\cal R}^{\pm,h,w}_{{\cal H},\overline {\cal H}}$
were identified with the operators ${\cal O}^{a,\overline
  a}_{n}$ in \cite{Pakman, Giribet:2007wp}.
To compare the corresponding three-point functions in the dual theories,
the two R-R-NS correlators needed are
\begin{equation}
{\mathbb A}_3
=g_s^{-2}\left\langle{\cal R}^{\pm,h_3,w_3}_{{\cal
    H}_3,\overline{\cal H}_3}(x_3,\overline x_3){\cal R}^{\pm,h_2,w_2}_{{\cal
    H}_2,\overline{\cal H}_2}(x_2,\overline x_2){\cal
    W}^{h_1,w_1}_{{\cal H}_1,\overline {\cal H}_1}(x_1,\overline x_1)
\right\rangle_{S^2}\, ,
\end{equation}
\begin{equation}
{\mathbb A}'_3=g_s^{-2}\left\langle{\cal
    Y}^{h_3,w_3}_{{\cal H}_3,\overline {\cal H}_3}(x_3,\overline x_3)
{\cal R}^{\pm,h_2,w_2}_{{\cal
    H}_2,\overline{\cal H}_2}(x_2,\overline x_2){\cal R}^{\pm,h_1,w_1}_{{\cal
    H}_1,\overline{\cal H}_1}(x_1,\overline x_1)
\right\rangle_{S^2}\, .
\end{equation}

The R vertices (\ref{vertex.ramonx}) were obtained in the $-\frac 12$
picture, so it is not necessary to insert a picture changing operator
and we can compute this amplitude for 
states in
arbitrary $w$ sectors, as long as $w_n=w_m+w_l$.

The SU(2) part of the three-point functions
 is given by $C'(j_i)$ for $j_n=j_m+j_l$ and the fermionic contributions
 are the following \cite{Giribet:2007wp}
\footnote{We get the inverse of the result
 reported
in \cite{Giribet:2007wp}.}
\begin{eqnarray}
\langle S^{w_3+\frac 12}(x_3,y_3)S^{w_2+\frac 12}
(x_2,y_2)\psi^{w_1+1}(x_1)\chi^{w_1}(y_1)
\rangle &=&\frac{(w_1+w_2)!}{w_1!w_2!}\, ,\nonumber\\
\langle \psi^{w_3}(x_3)\chi^{w_3+1}(y_3)S^{w_2+\frac 12}(x_2,y_2)
S^{w_1+\frac 12}(x_1,y_1)
\rangle
&=& \frac{(w_1+w_2)!}{w_1!w_2!}\, .
\end{eqnarray}

As shown in the previous section,
the SL(2,$\mathbb R)$ contribution is simply 
$C(h_i)$ for two or three flowed chiral primary states satisfying $m_1+m_2+m_3=0$.
If the three operators are flowed, the SU(2) spins
  violate the    triangular inequality and the
correlator
vanishes, analogously to the NS-NS-NS case.
When one operator is unflowed,  the factor
$\frac{(w_1+w_2)!}{w_1!w_2!}$ reduces to unity and we have
\begin{eqnarray}
{\mathbb A}_3 =
{\mathbb A}'_3 =
g_s^{-2}
\sqrt{B(h_1)
B(h_2)B(h_3)}\, .
\end{eqnarray}

Normalizing the R operators as (see \cite{Giribet:2007wp} for details)
\begin{equation}
{\mathbb R}^{\pm,h,w}_{{\cal H}, \overline {\cal
    H}}=\sqrt{\frac{(2h-1+kw)}{2 B(h)}}g_s {\cal R}^{\pm,h,w}_{{\cal H},
    \overline {\cal H}}\, ,
\end{equation}
we get
\begin{eqnarray}
\left\langle{\mathbb R}^{\pm,h_3,w_3}_{{\cal H}_3,\overline {\cal
    H}_3}{\mathbb R}^{\pm,h_2,w_2}_{{\cal H}_2,\overline {\cal H}_2}
    {\mathbb W}^{h_1,w_1}_{{\cal H}_1, \overline {\cal H}_1}
\right\rangle&=&
 \left\langle{\mathbb R}^{\pm,h_3,w_3}_{{\cal H}_3,\overline {\cal
    H}_3}{\mathbb R}^{\pm,h_2,w_2}_{{\cal H}_2,\overline {\cal H}_2}
    {\mathbb Y}^{h_1,w_1}_{{\cal H}_1, \overline {\cal H}_1}
\right\rangle\nonumber\\
&=&\frac {2g_s}{k^2}
\left
[\frac{(2h_3+kw_3-1)(2h_2+kw_2-1)}{(2h_1+kw_1-1)}\right ]^{1/2}\, ,~~~~~~
\label{rrw2}
\end{eqnarray}
for $w_1=0$ or $w_2=0$,
again in
agreement  with the boundary correlators (\ref{r1}) 
and (\ref{r2}).

\section{Conclusions}

We have evaluated spectral flow conserving 
three-point functions containing spectral flow images
 of chiral primary states in type IIB superstring theory on
AdS$_3\times$ S$^3\times$ T$^4$ and showed that they agree with the
corresponding correlators in the dual boundary CFT. These results
provide an additional verification of the AdS$_3$/CFT$_2$
correspondence, widening similar conclusions of previous works
\cite{Gaberdiel, Pakman, ps} to
 the non-trivial spectral flow sectors of the theory.

The matching obtained so far reflects the cancellation of the three-point 
structure
constant of AdS$_3$ against that of the
S$^3$ factor. 
The non-trivial fermionic contributions reduce to unity in
 all the non-vanishing
amplitudes 
that we have considered here. 
A definite confirmation of this duality
would require extending the bulk-to-boundary dictionary to
 descendant states. The evaluation of
three-point functions involving affine descendants and their spectral
flow images
 is an interesting subject in its own right. Actually, the spectral
flow operation maps primaries into descendants both in SU(2) and SL(2,
$\mathbb R)$ and 
it generates new representations of the universal cover of SL(2, $\mathbb R)$. 
Understanding these new representations is crucial to solve the
AdS$_3$ WZNW model
and elucidate the physical mechanism determining the truncation of the
 fusion rules imposed 
by the spectral flow symmetry \cite{wc}.
In the context of the AdS$_3$/CFT$_2$ correspondence, 
a better comprehension of the structure of the spectral
 flow sectors would contribute to achieve a systematic
 comprehension of the hypothesis advanced in the literature.

\bigskip

{\bf Acknowledgements:} We would like to thank
W. Baron, S. Iguri and P. Minces for useful discussions. We are
specially grateful to A. Pakman and an anonymous referee for 
carefully
reading the manuscript, for pointing out a
misleading conclusion in the previous version
of this paper  and for many interesting comments.
This work was supported in part by grants
PIP-CONICET/6332 and UBACyT X161.

\appendix

\section{Appendix: Clebsch-Gordan Coefficients}

In this Appendix we  compute the Clebsch-Gordan coefficients (CG)
expanding the product representation ($H\otimes\hat{h}$) of the
SL(2,$\mathbb R)$ algebra. We consider the case $H\in {\cal
D}^{+,w}_{H}$, $\hat{h}\in{\cal D}_{\hat{h}}$, where
\begin{equation}  \label{discretelowest} \mathcal{D}^+_{H}:\quad
\{|H,M\rangle\, ; ~~~ {H\in \mathbb R} ,~~~
 M=H+n,\quad n=1,2,3.....\}\, ,\end{equation} is an infinite
 discrete representation and
\begin{equation}  \label{discretespin} \mathcal{{D}}^+_{\hat{h}}:\quad
\{|\hat{h},\hat{m}\rangle\, ; \quad -\hat{h}\leq \hat{m}\leq
\hat{h}\, ,\quad \hat{m}\in \ {\mathbb Z} \} \, ,\end{equation} is
a finite representation of the SL(2,$\mathbb R)_{-2}$ algebra.
We use the following normalization
\begin{equation}
{\bf j}^\pm |H,M>=(M\mp H)|H, M\pm 1>
\end{equation}
and similarly for $|\hat h, \hat m>$.
A state living in the product representation may be expanded as
\begin{equation}  \label{clebschserie} |H\otimes
\hat{h}\rangle\equiv|H,\hat{h};\mathcal{H},\mathcal{\mathcal{M}}\rangle
=\sum_{M,\hat{m}}|H,M;\hat{h},\hat{m}\rangle\langle
H,M;\hat{h},\hat{m}|H,\hat{h};\mathcal{H},\mathcal{M}
\rangle\delta_{\mathcal{M},M+\hat{m}}\,
.
\end{equation}

Applying the raising operator $\bf{H}^+=\bf{j}^+_1+\bf{j}^+_2$ and
equating the coefficients on both sides  of  (\ref{clebschserie}),
the following recursion relation is  obtained
\begin{align}\label{relation1}
\begin{split}
 (\mathcal{M}-\mathcal{H})\langle \mathcal{M}+1-\hat{m},\hat{m}|\mathcal{H},\mathcal{M}+1\rangle&=(\mathcal{M}-\hat{m}-H)\langle
\mathcal{M}-\hat{m},\hat{m}|\mathcal{H},\mathcal{M}\rangle\\
&+(\hat{m}-1-\hat{h})\langle \mathcal{M}-\hat{m}
+1,\hat{m}-1|\mathcal{H},\mathcal{M}\rangle\, ,
\end{split}
\end{align}
where the indices $H,\ \hat{h}$ have been dropped for short.
 A similar
recursion relation is obtained applying the lowering operator
$\bf{H}^-=\bf{j}^-_1+\bf{j}^-_2$, namely
\begin{align}\label{relation2}
\begin{split} (\mathcal{M}+\mathcal{H})&\langle \mathcal{M}-1-\hat{m},
\hat{m}|\mathcal{H},\mathcal{M}-1\rangle=\\
&(\mathcal{M}-\hat{m}+H)\langle
\mathcal{M}-\hat{m},\hat{m}|\mathcal{H},\mathcal{M}\rangle
+(\hat{m}+\hat{h}+1)\langle
\mathcal{M}-\hat{m}-1,\hat{m}+1|\mathcal{H},\mathcal{M}\rangle\, .
\end{split}
\end{align}

The last term in (\ref{relation1}) vanishes for $\hat{m}=-\hat{h}$,
$i.e.$
\begin{equation}   \langle \mathcal{M}+\hat{h}+1,-\hat{h}|\mathcal{H},
\mathcal{M}+1\rangle =\frac{\mathcal{M}+\hat{h}-H}{\mathcal{M}-\mathcal{H}}
\langle
\mathcal{M}+\hat{h},-\hat{h}|\mathcal{H},\mathcal{M}\rangle\, ,
\end{equation} 
and for $\mathcal{M}=\mathcal{H}+1$, this reads
\begin{equation}   \langle \mathcal{H}+\hat h+2,-\hat{h}|\mathcal{H},
\mathcal{H}+2
\rangle=(\mathcal{H}+1+\hat{h}-H)\langle
M',-\hat{h}|\mathcal{H}, \mathcal{H}+1\rangle\, .
\end{equation} Then, taking successively $\mathcal{M}=\mathcal{H}+2,\cdots , 
\mathcal{H}+n$, one
finds
\begin{equation}  \label{m1funcj} \langle
M,-\hat{h}|\mathcal{H},\mathcal{M}\rangle=\frac{(\hat{h}-H+\mathcal{M}-1)!}
{(\mathcal{M}-\mathcal{H}-1)!(\mathcal{H}+\hat{h}-H)!}\langle
M'_1,-\hat{h}|\mathcal{H}, \mathcal{H}+1\rangle .\end{equation}

Defining
 $q(\hat{m},\mathcal{M})\equiv
\frac{(-1)^{\hat{m}}(\hat{m}+\hat{h})!(\mathcal{M}-\hat{m}+H)!}
{(\mathcal{M}+\mathcal{H})!}$,
 (\ref{relation2}) may be recast as
\begin{eqnarray}
 q(\hat{m}+1,\mathcal{M})\langle
\mathcal{M}-(\hat{m}+1),\hat{m}+1|\mathcal{H},\mathcal{M}\rangle&=&
q(\hat{m},\mathcal{M})\langle \mathcal{M}-\hat{m},\hat{m}|\mathcal{H},\mathcal{M}\rangle\nonumber\\
&&-q(\hat{m},\mathcal{M}-1)\langle
\mathcal{M}-1-\hat{m},\hat{m}|\mathcal{H},\mathcal{M}-1\rangle\nonumber\\
&\equiv&\Delta_\mathcal{M} [q(\hat{m},\mathcal{M})\langle
M',\hat{m}|\mathcal{H},\mathcal{M}\rangle]\, .
\end{eqnarray}
Applying this  successively for $\hat{m}-1,\cdots,\hat{m}-n$
and using
$ \Delta_x^n[f](x)=\sum_{s=0}^{n}(-1)^s\begin{pmatrix}
n\\
s\end{pmatrix}f(x-s)$, we get
\begin{eqnarray}
&&\langle
\mathcal{M}-\hat{m},\hat{m}|\mathcal{H},\mathcal{M}\rangle=\frac
1{q(\hat{m},\mathcal{M})}\Delta_\mathcal{M}^{\hat{m}+\hat{h}}
[q(-\hat{h},\mathcal{M})\langle M',-\hat{h}|\mathcal{H},\mathcal{M}
\rangle]\nonumber\\
&&=\frac
1{q(\hat{m},\mathcal{M})}\sum_{s=0}^{\hat{m}+\hat{h}}(-1)^s
\begin{pmatrix}\hat{m}+\hat{h}\\s\end{pmatrix}
q(-\hat{h},\mathcal{M}-s)\langle
\mathcal{M}-s+\hat{h},-\hat{h}|\mathcal{H},\mathcal{M}-s\rangle\, . 
\nonumber
\end{eqnarray}

Substituting $q(\hat{m},\mathcal{M})$ and $\langle
\mathcal{M}-s+\hat{h},-\hat{h}|\mathcal{H},\mathcal{M}-s\rangle$
 in this equation,
we  obtain
\begin{align}
\begin{split}
\langle \mathcal{M}-\hat{m},\hat{m}|\mathcal{H},\mathcal{M}\rangle&=
\frac{(\mathcal{M}+\mathcal{H})!}{(\hat{m}+\hat{h})!(\mathcal{M}-\hat{m}+H)!}
\sum_{s=0}^{\hat{m}+\hat{h}}(-1)^{s-\hat{h}}
\begin{pmatrix}\hat{m}+\hat{h}\\s\end{pmatrix}\frac{(\mathcal{M}-s+\hat{h}+H)!}
{(\mathcal{M}-s+\mathcal{H})!}\\
&\times\frac{(\hat{h}-H+\mathcal{M}-s-1)!}{(\mathcal{M}-s-\mathcal{H}-1)!
(\mathcal{H}+\hat{h}-H)!}\langle
\mathcal{H}+1+\hat{h},-\hat{h}|\mathcal{H},\mathcal{H}+1\rangle\,
.\nonumber
\end{split}
\end{align}

Therefore, all the CG coefficients in the expansion
(\ref{clebschserie}) are expressed 
in terms of just one coefficient, which  can be
set to one \footnote{ Recall that  the CG are  determined up
to a global phase factor (which is a global multiplicative factor
for all remaining CG).}. As a consistency check, we 
compute some known cases.

In the unflowed sector, we need to decompose the product
representation with $\hat{h}=1$. In this case, there are three
possible combinations of $\mathcal{H}$, according to the angular
momentum selection rules, namely
$
 \mathcal{H}=H+1,~\mathcal{H}=H,~ \mathcal{H}=H-1$.
In the first case, one gets
\begin{eqnarray} (\psi\Phi)^{\om=0}_{H+1,M}&=&
\sum_{M,\hat{m}}(\Phi^{\om=0}_{H,M}\psi^{\hat{m}})\langle
M-\hat{m},\hat{m}|H+1,M\rangle\nonumber\\
&=&\frac{1}{2}(H-M)(1+H-M)\Phi^{\om=0}_{H,M+1}\psi^{-}+(H+1-M)(1+H+M)
\Phi^{\om=0}_{H,M}\psi^{3}\nonumber\\
&+&\frac{1}{2}(H+M)(1+H+M)\Phi^{\om=0}_{H,M-1}\psi^{+} \,
.\end{eqnarray}

For $\mathcal{H}=H$, 
the following field expansion is obtained \ber\label{expanssionphi1}
(\psi\Phi)^{\om=0}_{H,M}&=&
(M-H)\Phi^{\om=0}_{H,M+1}\psi^{-}-2M\Phi^{\om=0}_{H,M}\psi^{3}+(H+M)
\Phi^{\om=0}_{H,M-1}\psi^{+}\, . \eer

And finally, for $\mathcal{H}=H-1$, which satisfies the chirality
condition in the unflowed sector,
\ber (\psi\Phi)^{\om=0}_{H-1,M}
&=&\Phi^{\om=0}_{H,M+1}\psi^{-}-2\Phi^{\om=0}_{H,M}\psi^{3}+
\Phi^{\om=0}_{H,M-1}
\psi^{+} \, ,\eer in accord with  the decomposition given
in \cite{Kutasov98}, up to the global phase factor mentioned above.

\end{document}